\documentclass{aa}
\usepackage{graphicx}
\usepackage{ulem}

      \def\new#1 {{\bf #1 }}
      \def\cut#1 {\sout{#1} }

\begin{document}
\def\ffam {\hbox{$\,.\!\!^{\prime}$}}
\def\ffas {\hbox{$\,.\!\!^{\prime\prime}$}}
\def\ffM {\hbox{$\,.\!\!^{\rm M}$}}
\def\ffm {\hbox{$\,.\!\!^{\rm m}$}}
\def\ffs {\hbox{$\,.\!\!^{\rm s}$}}

\title{The kinetic temperature of a molecular cloud at redshift 0.9:
       Ammonia in the gravitational lens PKS\,1830--211}


\author{C.~Henkel\inst{1}
        \and
        J.A.~Braatz\inst{2}
        \and
        K. M.~Menten\inst{1}
        \and
        J. Ott\inst{2,3}}

\offprints{C. Henkel, \email{chenkel@mpifr-bonn.mpg.de}}

\institute{Max-Planck-Institut f\"ur Radioastronomie, Auf dem H\"ugel 69, D-53121 Bonn, Germany
           \and
           National Radio Astronomy Observatory, 520 Edgemont Rd., Charlottesville, VA 22903, USA
           \and
           CSIRO, Australia Telescope National Facility, Cnr. Vimiera \& Pembroke Roads, Marsfield, NSW 2122, Australia}

\date{Received date / Accepted date}

\abstract
{Little is known about the structure of the interstellar medium and the nature of individual clouds in galaxies at 
intermediate redshifts. The gravitational lens toward PKS\,1830--211 offers the unique possibility to study this
interstellar gas with high sensitivity and angular resolution in a molecular cloud that existed half a Hubble time ago.}
{This multi-line study aims at a better definition of the physical properties of a significantly redshifted cloud.}
{Using the Green Bank Telescope (GBT), we searched for ammonia (NH$_3$) toward PKS\,1830--211.}
{We have detected all ten observed metastable $\lambda$$\sim$2\,cm ammonia lines. The ($J$,$K$) = (1,1) to (10,10) transitions,
up to $\sim$1030\,K above the ground state, were measured in absorption against the radio continuum of the lensed background 
source. The ammonia absorption appears to be optically thin, with absolute peak flux densities up to 2.5\% of the total continuum 
flux density. Measured intensities are consistent with a kinetic temperature of $\sim$80\,K for 80--90\% of the ammonia 
column. The remaining 10--20\% are warmer, with at least some of this gas reaching kinetic temperatures of $\ga$600\,K. 
Toward the south-western continuum source, the column density is $N$(NH$_3$) $\sim$ (5--10)$\times$10$^{14}$\,cm$^{-2}$, 
which implies a fractional abundance of $X$(NH$_3$) $\sim$ (1.5--3.0)$\times$10$^{-8}$. Similarities with a hot NH$_3$ 
absorption component toward the Sgr~B2 region close to our Galactic center, observed up to the (18,18) line, suggest that 
the Sgr~B2 component also consists of warm diffuse low-density gas. The warm absorption features from PKS\,1830--211 are 
unique in the sense that they originate in a spiral arm.}
{}

\titlerunning{Molecular absorption lines in PKS\,1830--211}

\authorrunning{Henkel et al.}

\maketitle


\section{Introduction}

Studies of interstellar emission and absorption lines are complementary. While emission commonly traces an extended gas 
reservoir, which provides information on global properties such as mass and kinematics, absorption tends to trace a spatially far
smaller region confined by the angular extent of a continuum background source. Absorption lines allow us to derive optical 
depths directly from line to continuum flux density ratios; the strength of the absorption depends on the flux of the background
source and is decoupled from the distance to the absorber. Thus, for objects at significant redshifts, studies of interstellar
absorption lines have the potential to provide unique information on otherwise inaccessible regions, combining extremely
high sensitivity with outstanding spatial resolution (e.g., Chengalur et al. \cite{chengalur99}; Wiklind \& Combes \cite{wc99}).

In spite of extended surveys, multi-line molecular absorption systems have so far only been discovered toward five distant 
targets (Wiklind \& Combes \cite{wc94}, \cite{wc95}, \cite{wc96a}; \cite{wc96b}; Kanekar et al. \cite{kanekar05}), all of 
which are located at intermediate redshifts (0.25$\leq$$z$$\leq$0.89). The two most notable absorbers, those toward B0218+357 
and PKS\,1830--211, share a number of common properties. They have similar redshifts ($z$=0.68 and 0.89, respectively), show 
the highest line-of-sight column densities, and originate in gravitational lenses. Toward both systems the respective background 
sources, probably a BL Lac object and a blazar, respectively (e.g. Kemball et al. \cite{kem01}; De Rosa et al. \cite{ros05}), 
are lensed into three dominant features, a north-eastern and a south-western hotspot and an Einstein ring. Optical or near 
infrared and radio separations of the hotspots differ due to optical obscuration (Courbin et al. \cite{courbin98}; York et al. 
\cite{york05}). The lensing galaxies themselves are spirals viewed almost face-on (Courbin et al. \cite{courbin02}; Winn et al. 
\cite{winn02}; York et al. \cite{york05}). The bulk of the absorption arises in both cases from in front of the south-western of 
the two main radio continuum images, at radial distances of $\sim$2 and $\sim$4\,kpc from the center of the respective lens (e.g., 
Frye et al. \cite{frye97}; Swift et al. \cite{swift01}; Meylan et al. \cite{mey05}; York et al. \cite{york05}). In both cases, 
this absorption originates in the south-western spiral arm of the parent galaxy.

There are also significant differences between B0218+357 and PKS\,1830--211: one is image separation (334 versus 970\,mas; e.g., 
Jin et al. \cite{jin03}; Wucknitz et al. \cite{wuck04}); another is time delay (10 versus 25 days; e.g., Lovell et al. 
\cite{lov98}; Biggs et al. \cite{biggs01}). With respect to the detection of molecular absorption lines, the most striking 
differences are the continuum levels and the complexity of the absorption systems. B0218+357 hosts two molecular components 
separated by $\sim$13\,km\,s$^{-1}$ observed toward the south-western continuum image (Menten \& Reid \cite{menten96}; Jethava 
et al. \cite{jethava07}; Muller et al. \cite{muller07}). Toward PKS\,1830--211, weak molecular absorption is also measured 
toward the north-eastern radio continuum hotspot, displaced by \hbox{--147}\,km\,s$^{-1}$ from the dominant south-western 
absorption component (Wiklind \& Combes \cite{wc98}). Furthermore, there is H{\sc i} absorption at $z$=0.19 (Lovell et al. 
\cite{lovell96}).

PKS\,1830--211 is one of the most prominent compact radio sources in the sky and therefore an ideal target for absorption line 
studies at radio wavelengths. Molecular species that have been detected include CS, HCN, HCO$^+$, HNC and N$_2$H$^+$ (Wiklind
\& Combes \cite{wc96b}), CO (Gerin et al. \cite{gerin97}), OH (Chengalur et al. \cite{chengalur99}) and C$_2$H, H$_2$CO,
C$_3$H$_2$, and HC$_3$N (Menten et al. \cite{menten99}). The $\lambda$21\,cm line of H{\sc i} was also observed at
$z$=0.89 (Chengalur et al. \cite{chengalur99}). Muller et al. (\cite{muller06}) determined CNO and sulfur isotope ratios.

PKS\,1830--211 provides a unique view of a molecular cloud at more than half a Hubble time in the past ($\Lambda$-cosmology 
with $H_{0}$=73\,km\,s$^{-1}$\,Mpc$^{-1}$, $\Omega_{\rm m}$ = 0.28 and $\Omega_{\Lambda}$ = 0.72; Spergel et al. \cite{sper07}; 
1\,mas corresponds to $\sim$7.5\,pc). With a luminosity distance of $\sim$5.5\,Gpc it is the farthest molecular cloud known to 
date that can be studied in detail. The nature of this object is, however, poorly understood. In an attempt to constrain the 
physical properties of the target and motivated by the recent detection of NH$_3$ toward B0218+357 (Henkel et al. \cite{henkel05}), 
we searched for NH$_3$ absorption to evaluate the still poorly constrained physical parameters of the absorbing molecular gas.

\section{Observations}

The ($J$,$K$) = (1,1) to (7,7) inversion lines of NH$_3$ were observed with the Green Bank Telescope (GBT) of the
NRAO\footnote{The National Radio Astronomy Observatory is a facility of the National Science Foundation operated under
cooperative agreement with Associated Universities, Inc.} on 2003 April 24. On 2004 May 21--22, the (1,1), (2,2), (4,4)
to (6,6), and (8,8) to (10,10) lines were also measured. The lines observed with a beam size of $\sim$1\arcmin\ cover 
a frequency range of 12.56--15.17\,GHz, accounting for a redshift of $z$=0.88582. We used a dual-beam Ku-band receiver 
in a total power observing mode. The source was positioned alternately between the two beams, cycling after 2 minutes 
at each position. During the 2003 observations, we employed an electronic beam switch. Because we found that the beam 
switch did not improve baseline performance, it was not used during the 2004 observations.

During the 2003 observations, we configured the backend in a single spectral window covering 50\,MHz. The (1,1) and
(2,2) lines fit in a single 50 MHz window and were observed simultaneously, while all other lines were measured
individually. During the 2004 observations, two spectral windows were measured simultaneously, each window centered
on a line. We determined the continuum level of PKS\,1830--211 using the GBT continuum backend during pointing scans.
We first calibrated the flux density scale using measurements of the flux calibrators 3C286 and 3C48, and then adjusted this
scale for elevation-dependent gain variations.

Data were processed using AIPS++. We determined a mean continuum level from the baseline of each spectrum and then scaled 
the spectra to match this continuum level to the value measured during the pointing scans. For spectral line analysis, a low 
order ($\leq$2) polynomial baseline was fit and removed from the unsmoothed spectra. We estimate that the overall uncertainty 
in the flux density scale of the absorption lines is $\pm$15\%. The (1,1) and (2,2) spectra were affected by interference that 
caused a ripple in the baseline and contributed additional uncertainty to the flux, estimated to be $\pm$10\%.

\begin{figure}[t]
\vspace{-0.3cm}
\centering
\resizebox{8.6cm}{!}{\rotatebox[origin=br]{0}{\includegraphics{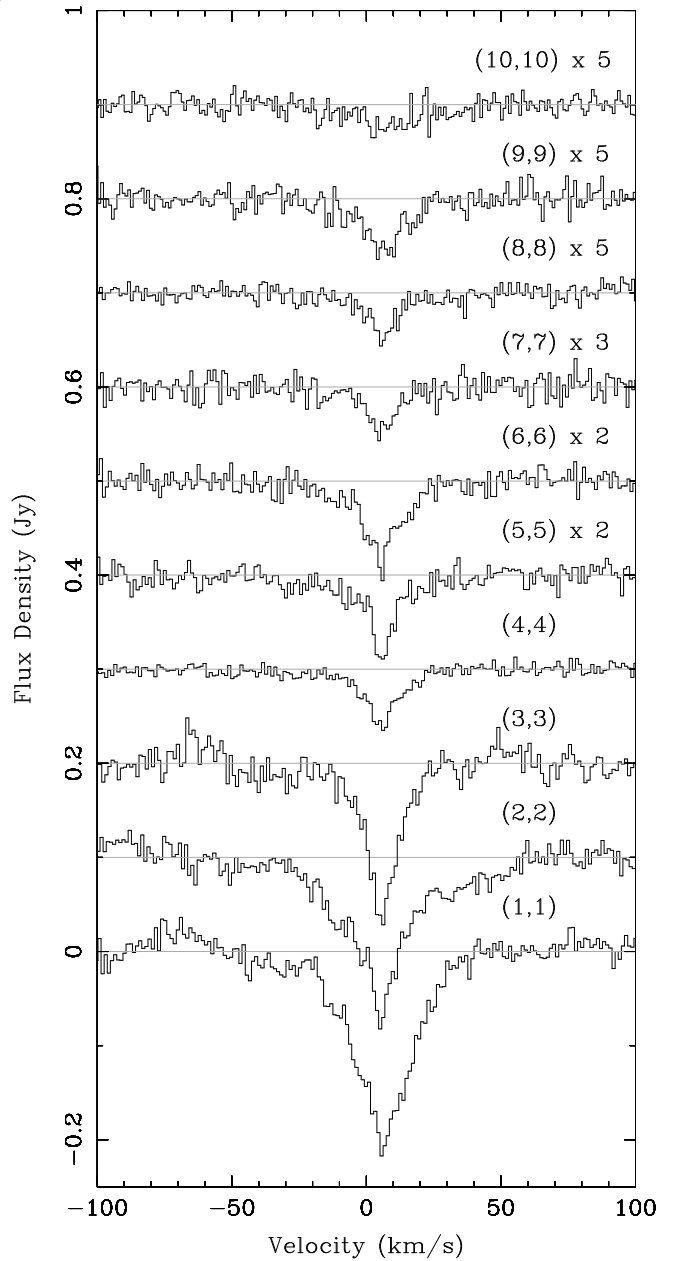}}}
\vspace{-0.1cm}
\caption{NH$_3$ lines with a Local Standard of Rest (LSR) velocity scale relative to $z$=0.88582, observed toward PKS\,1830--211.
The ($J$,$K$) = (1,1) to (7,7) lines were observed in 2003, the (8,8) to (10,10) lines represent data from 2004. The channel spacing 
is 1.1\,km\,s$^{-1}$. Satellite groups of hyperfine components, shifted by about $\pm$10 and $\pm$20\,km\,s$^{-1}$ relative to the
main component, may be visible in the (1,1) and (2,2) profiles.
\label{fig1}}
\end{figure}

\section{Results}

The continuum flux density of PKS\,1830--211 was found to be 8.12\,Jy at 12.8\,GHz on 2003 April 24 and 6.45\,Jy at 14.0\,GHz 
on 2004 May 21. We estimate the uncertainty to be $\pm$10\%. Fig.\,\ref{fig1} shows the measured ammonia (NH$_3$) absorption profiles 
up to the ($J,K$)=(10,10) line, $\sim$1030\,K above the ground state. Obviously, the presence of lines from such highly excited 
states requires the presence of a warm molecular medium. Absorption displaced by \hbox{--147}\,km\,s$^{-1}$ was not seen. While 
baseline ripples prevent the determination of a reliable upper limit for the (1,1) line of this velocity component, 1$\sigma$ 
noise levels of the (2,2) and (3,3) lines are 9 and 10\,mJy for a 1.1\,km\,s$^{-1}$ channel, respectively.

The inversion lines are characterized by a narrow main feature of width $\sim$15\,km\,s$^{-1}$, centered at slightly positive
velocities and reaching in the (1,1) line about --2.5\% of the total continuum flux density. In the (1,1), (2,2), and perhaps the 
(3,3) line there is also a wider component that can be interpreted as evidence for satellite absorption at offsets of about 
$\pm$10 and $\pm$20\,km\,s$^{-1}$ from the main feature (see e.g. Ho \& Townes \cite{ho83} for a detailed (1,1) spectrum). 
Alternatively, the blue-shifted line wings may show the --21 and --3\,km\,s$^{-1}$ features mentioned by Muller et al. 
(\cite{muller06}). Irrespective of their origin, the weakness of these absorption components indicates optically thin absorption, 
which is consistent with single-component Gaussian fits that incorporate the hyperfine structure of the (1,1) to (3,3) lines.

Table 1 displays integrated optical depths, velocities, and linewidths for all measured transitions. Optical depths were derived 
for a source covering factor $f_{\rm c}$ = 1 using
$$
\tau = -{\rm ln} \left( \frac{1 - |T_{\rm L}|}{T_{\rm c}}\right).
$$
with $T_{\rm L}$ and $T_{\rm c}$ denoting line and continuum temperature, respectively.

Comparing the data from 2003 and 2004 (this refers to the (1,1), (2,2), and (4,4) to (6,6) lines), the perfect agreement of
integrated optical depths in the (5,5) line is fortuitous, but the agreement is generally very good. This indicates that line and 
continuum flux densities are well correlated. For those transitions that were observed twice, the 2003 data tend to have higher 
signal-to-noise ratios and are therefore shown in Fig.\,\ref{fig1}.

\begin{figure}[t]
\vspace{-0.3cm}
\centering
\resizebox{!}{7.2cm}{\rotatebox[origin=br]{-90}{\includegraphics{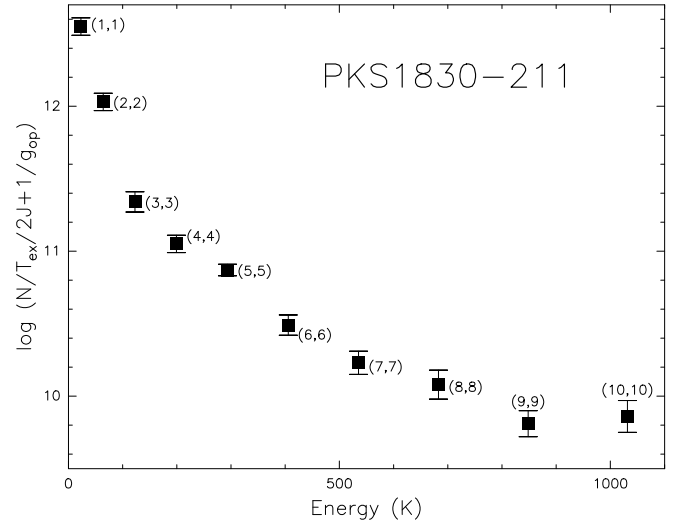}}}
\caption{Normalized total column densities of NH$_3$ inversion doublets versus energy above the ground state (see
Sect.\,3) for a source covering factor $f_{\rm c}$ = 1. For those lines that were measured twice, a weighted 
average was calculated.
\label{fig2}}
\end{figure}

Figure~\ref{fig2} shows a ``Boltzmann plot'' (rotation diagram) including the ten measured ``metastable'' ($J$=$K$) NH$_3$
inversion lines. The sum of the column densities of the two states of an inversion doublet,
$$
N(J,K)/T_{\rm ex}  = 1.65\times 10^{14}\,\,\,\,\frac{J(J+1)}{K^2\,\,\nu} \int{\tau\,\,{\rm d}V}
$$
($N(J,K)$ in cm$^{-2}$; $T_{\rm ex}$ in K; $\nu$ in GHz), is normalized and plotted against the energy above the ground state. 
Normalization is obtained by dividing $N(J,K)$ by the statistical weight of the respective transition, which is (2$J$+1)\,$g_{\rm op}$, 
with $g_{\rm op}$ = 1 for para-NH$_3$ ($K$ = 1, 2, 4, 5, 7, 8, 10) and $g_{\rm op}$ = 2 for ortho-NH$_3$ ($K$=3, 6, 9). 
The a priori unknown excitation temperature of the observed inversion lines is denoted by $T_{\rm ex}$. The levels of a given
inversion doublet are separated by only $\sim$1.2\,K. However, depending on their quantum numbers $J$ and $K$, these doublets 
cover an enormous range of excitation conditions.

Since critical densities are similar (due to similar frequencies, Einstein coefficients, and collision rates; e.g., Danby et al.
\cite{danby88}; {\it http://www.strw.leidenuniv.nl/$\sim$moldata/}), we can assume that the excitation temperature $T_{\rm ex}$
is the same for all observed metastable inversion lines. If all the level populations were determined by a single rotational 
temperature, the distribution of data points in Fig.\,\ref{fig2} would be well fitted  by a straight line whose inverse slope 
would be a measure of that temperature, $T_{\rm rot}$. Since radiative transitions between different metastable inversion 
doublets are, to first order, forbidden and since the metastable doublets are located at the bottom of their respective 
``$K$-ladder'' (see Rohlfs \& Wilson \cite{rohlfs03} for a level diagram), the population of the metastable levels cannot 
rapidly decay to lower states and $T_{\rm rot}$ is also a measure of kinetic temperature.

From Fig.\,\ref{fig2}, rotation temperatures can at best be fitted in a piecemeal fashion in the following sense: For para-ammonia, 
a comparison of the ($J$,$K$) = (1,1) and (2,2) lines gives $T_{\rm rot}$$\sim$35\,K. For the (2,2) and (4,4) lines, the 
corresponding value becomes $\sim$60\,K. The (4,4) and (7,7) lines imply $\sim$175\,K and the (7,7) to (10,10) lines indicate 
$\sim$600\,K. The ortho-ammonia lines fit well into this general trend of steeply rising rotation temperature with increasing 
rotational quantum number $J$. Since frequencies of the ammonia lines are quite similar (Sect.\,2), the resulting temperatures 
are not strongly affected by differences with respect to beam size or morphology of the background continuum.

Adding up the column densities of the observed inversion doublets, we obtain $N$(NH$_3$)/$T_{\rm ex}$ = 2.3$\times$10$^{13}$\,cm$^{-2}$/K,
with 4.1$\times$10$^{12}$\,cm$^{-2}$/K of this column belonging to the ortho-species. The populations of the (1,1) to (3,3) 
doublets cover 85\% of the total observed column density, while the (10,10) doublet contributes 0.7\% (see Sect.\,4.3 for a 
more detailed discussion).

\begin{table}
\caption{NH$_3$ line parameters$^{\rm a)}$}
\begin{center}
\begin{tabular}{l c r r @{\ } r}
\hline
Line          & $\int{\tau {\rm d}V}$ & $V$\ \ \ \ \ \ \  & \multicolumn{2}{c}{$\Delta V_{1/2}$} \\
              & \multicolumn{4}{c}{(km\,s$^{-1}$)}                        \\
\hline
NH$_3$ (1,1)  & 0.661\,(0.120)         &  +9.0\,(0.1) & 27.3&(0.3)        \\
              & 0.895\,(0.164)         &  +8.6\,(0.5) & 34.9&(1.2)        \\
NH$_3$ (2,2)  & 0.599\,(0.109)         &  +7.4\,(0.2) & 33.6&(0.5)        \\
              & 0.429\,(0.082)         & +10.5\,(0.4) & 20.8&(1.0)        \\
NH$_3$ (3,3)  & 0.330\,(0.050)         &  +8.6\,(0.1) & 15.6&(0.3)        \\
NH$_3$ (4,4)  & 0.137\,(0.021)         &  +8.4\,(0.2) & 17.3&(0.4)        \\
              & 0.105\,(0.022)         &  +9.1\,(0.7) & 14.7&(1.7)        \\
NH$_3$ (5,5)  & 0.101\,(0.016)         &  +8.3\,(0.3) & 16.9&(0.6)        \\
              & 0.101\,(0.017)         &  +8.7\,(0.5) & 20.6&(1.3)        \\
NH$_3$ (6,6)  & 0.111\,(0.017)         &  +8.1\,(0.2) & 20.4&(0.6)        \\
              & 0.091\,(0.040)         &  +8.5\,(0.6) & 14.8&(1.4)        \\
NH$_3$ (7,7)  & 0.035\,(0.006)         & +10.9\,(0.4) & 13.4&(0.9)        \\
NH$_3$ (8,8)  & 0.029\,(0.006)         &  +8.5\,(0.6) & 14.8&(1.4)        \\
NH$_3$ (9,9)  & 0.037\,(0.007)         &  +8.8\,(0.7) & 20.7&(1.6)        \\
NH$_3$ (10,10)& 0.024\,(0.005)         & +10.2\,(2.0) & 36.9&(5.0)        \\
\hline
\end{tabular}
\end{center}
a) From Gaussian fits adopting $z$=0.88582 and assuming that the continuum source covering factor is $f_{\rm c}$=1.
Standard deviations are given in parentheses. In cases in which a line was measured twice, the upper line refers to data
taken in 2003 (see Sect.\,2). The errors in the integrated optical depths include a 15\% calibration uncertainty,
a 10\% uncertainty due to interference in the (1,1) and (2,2) spectra, and the standard deviation from the Gaussian
fit. For $V$ and $\Delta V_{1/2}$, standard deviations were obtained from Gaussian fits alone.
\end{table}

\section{Discussion}

\subsection{The nature of the absorbing cloud}

The molecular lens in PKS\,1830--211 was originally considered to host a diffuse cloud, possibly corresponding to the
intercloud medium of a giant molecular complex (Wiklind \& Combes \cite{wc96b}). Subsequently, the chemistry of the
absorbing cloud was found to show a number of properties reminiscent of Galactic cold dark clouds (Menten et al.
\cite{menten99}; Shah et al. \cite{shah99}; Muller et al. \cite{muller06}). Menten et al. (\cite{menten99}) found
remarkable similarities to the dark cloud TMC-1, apart from C$_3$H$_2$ and HC$_3$N, which are known to be anomalously 
abundant in TMC-1. For many molecules, including C$_3$H$_2$ and HC$_3$N, but excluding CS and C$_2$H, there is also good 
agreement with measurements of the dark cloud L\,183 (L\,134\,N). Shah et al. (\cite{shah99}) mention that there is little 
evidence that physical conditions vary substantially between the PKS\,1830--211 column and the column of the dark cloud 
L\,134. They note, however, that the deuterium enhancement of PKS\,1830--211 is too low for a ``normal'' dark cloud, a 
conclusion that is based on their non-detections of DCN and DCO$^+$. Muller et al. (\cite{muller06}) find a [HCN]/[HNC] 
abundance ratio of $\sim$2.5, which they interprete as support for the dark cloud scenario even though the ratio is at 
the very upper end of the range of values for this class of objects (e.g., Hirota et al. \cite{hirota98}). Muller et al. 
(\cite{muller06}) also report the detection of an anomalously high abundance of H$_2$S in PKS\,1830--211. They suggest 
that this may be either a consequence of a high sulfur abundance or a sign of stellar activity. In the latter case, H$_2$S 
would be enhanced due to UV irradiation of dust grains by young massive stars or due to shocks (e.g., Codella et al. 
\cite{codella05}; Jim{\'e}nez-Serra et al. \cite{jimenez05}).

Our detection of highly excited ammonia demonstrates that {\it the dark cloud scenario does not hold and that there must be
some kind of activity enhancing kinetic temperatures to high values}. In view of a considerable distance to the center of
the absorbing galaxy ($\sim$4\,kpc, see Sect.\,1), the high temperatures may either be triggered by the radiation of newly
formed massive stars or by large-scale shocks (see, e.g., Flower et al. \cite{flower95} and Rizzo et al. \cite{rizzo01} 
for NH$_3$ in a shocked environment). This may explain the high H$_2$S abundance reported by Muller et al. (\cite{muller06}). 
A high kinetic temperature also reduces isotope fractionation so that the non-detection of deuterated species (Shah et al. 
\cite{shah99}) is no surprise.

\subsection{Kinetic temperatures}

The concave shape of the line connecting points in the Boltzmann diagram (Fig.\,\ref{fig2}) indicates rising rotation
temperatures with increasing excitation above the ground state. This is similar to results from Galactic sources (e.g.,
Wilson et al. \cite{wilson93, wilson06}) and was originally interpreted to indicate a gradient in $T_{\rm kin}$, possibly
suggesting the presence of dense post-shock gas that is gradually cooling with increasing distance from the shock front.
We note, however, that in a warm environment radiative transfer calculations predict higher $T_{\rm rot}$ values for higher
metastable levels, gradually approaching $T_{\rm kin}$, even if the gas is characterized by a single kinetic temperature
(Walmsley \& Ungerechts \cite{walmsley83}; Danby et al. \cite{danby88}; Flower et al. \cite{flower95}). $T_{\rm rot}$ only
provides lower limits to $T_{\rm kin}$. A major source of uncertainty in the interpretation of Fig.\,\ref{fig2} is caused
by the fact that collision rates (Danby et al. \cite{danby88}) were only computed for levels up to ($J$,$K$)=(6,6) and
temperatures $\leq$300\,K. For higher transitions and temperatures, rate coefficients have to be extrapolated.

Such an extrapolation, even including vibrationally excited states, was carried out by Schilke (\cite{schilke89}). Adopting
his results, assuming optically thin lines (a realistic assumption, see Sect.\,3), and ignoring any infrared radiation field, 
a kinetic temperature of $T_{\rm kin}$ $\sim$ 80\,K reproduces well the measured ($J$,$K$) = (1,1), (2,2), and (4,4) rotation 
temperatures. However, the ($J$,$K$) $>$ (4,4) transitions are far too strong and require much higher excitation. To summarize, 
80--90\% of the observed NH$_3$ column originates in gas with $T_{\rm kin}$ $\sim$ 80\,K. In addition to this dominant 
component, there is warmer gas, which is characterized by kinetic temperatures well in excess of 100\,K. This hot component 
dominates the ($J$,$K$) $>$ (4,4) metastable lines.

\subsection{The ammonia column density}

At the end of Sect.\,3, we derived NH$_3$ column densities for the observed metastable inversion doublets. While metastable
doublets beyond ($J$,$K$) = (10,10) should not play an important role, we have not yet estimated the continuum source 
covering factor, and have also ignored contributions from the ($J$,$K$) = (0,0) ground state and from the non-metatastable 
levels ($J$$>$$K$).

The absorbed south-western continuum source of PKS\,1830--211 (see Sect.\,1) contributes to $\sim$40\% of the  total radio
flux density (e.g., Nair et al. \cite{nair93}; van Ommen et al. \cite{ommen95}). According to Frye et al. (\cite{frye97}),
its molecular cloud coverage is $\sim$70\%. This indicates a total source covering factor of order $f_{\rm c}$ $\sim$ 0.3, 
which should be taken with some scepticism because the coverage was determined at 94\,GHz and not at 12--15\,GHz. At frequencies
approximately equal to those of the ammonia lines reported here, Menten et al. (\cite{menten99}) find a source size 
$\ga$2.5\,mas for HC$_3$N absorption. This appears to be similar to the size of the continuum source (e.g., Garrett et al. 
\cite{garrett97}) and also suggests that the molecular coverage of the south-western continuum source alone is not far 
below unity.

To estimate the population of the ($J$,$K$) = (0,0) ground state, we should keep in mind that if ammonia is synthesized and
equilibrated at high-temperatures ($\ga$40\,K), the ortho-to-para abundance ratio should be close to one. If, however, this 
temperature is lower, the ratio should be larger because the lowest state, ($J$,$K$) = (0,0), belongs to the ortho-species (for
the extreme case of 5\,K, this would yield an ortho-to-para abundance ratio in excess of 10). Since one half of the levels in
the $K$=0 ladder are missing, the (0,0) level does not form a doublet. For an ortho-to-para ratio of unity and accounting for 
the measured inversion lines, a rotation temperature of 30\,K is required between the (0,0) and (3,3) states. This is consistent 
with $T_{\rm rot}$ $\sim$ 35\,K from the (1,1) and (2,2) inversion doublets, which are located at intermediate energies above 
the ground state, and we obtain $N$(NH$_3$)/$T_{\rm ex}$ $\sim$ $f_{\rm c}^{-1}$ $\times$ 3.5$\times$10$^{13}$\,cm$^{-2}$/K 
$\sim$ 10$^{14}$\,cm$^{-2}$/K.

To populate the non-metastable ($J$$>$$K$) levels, either high spatial densities or strong radiation fields are needed
(e.g., Mauersberger et al. \cite{mauers85}). Toward PKS\,1830--211, however, densities must be low because excitation
temperatures of molecular mm-wave transitions are close to the temperature of the cosmic microwave background at $z$=0.886, 
$T_{\rm CMB}$ = 5.15\,K (see Sect.\,4.4). Radiative excitation of the rotational transitions of ammonia ($\Delta$$J$ = 1, 
$\Delta$$K$ = 0) at far infrared wavelengths ($\sim$50--200$\mu$m; see Ceccarelli et al. \cite{ceccarelli02}) is also not a 
viable option. It would require a cloud with high column density ($N$(H$_2$) $>$ 10$^{23}$\,cm$^{-2}$) that is embedded in 
warm dust ($T_{\rm dust}$ $\ga$ 100\,K) to avoid geometrical dilution and a weakening of the radiation field due to 
insufficient optical depths. In the Galaxy, such regions containing large column densities of warm dust are always 
associated with dense ($n$(H$_2$)$>$10$^4$\,cm$^{-2}$) molecular gas. However, we do not see such a gas component toward 
PKS\,1830--211. In view of these arguments, $N$(NH$_3$)/$T_{\rm ex}$ = 10$^{14}$\,cm$^{-2}$/K is the most likely value 
because it is based on an ortho-to-para abundance ratio of one, neglecting potentially small contributions from the 
non-metastable levels. In view of the poorly constrained source covering factor, density, and ortho-to-para ratio of the 
cloud, we cannot assign an uncertainty to this $N$(NH$_3$)/$T_{\rm ex}$ estimate.

\subsection{Chemical considerations}

In Table~1 of Menten et al. (\cite{menten99}) and Table~8 of Muller et al. (\cite{muller06}), column densities of various 
molecular species are listed. With $N$(CO) = 3$\times$10$^{18}$\,cm$^{-2}$ and $N$(H$_2$)/$N$(CO) = 10$^{4}$ (for the latter, 
see, e.g., Mathur \& Nair \cite{mathur97}; Wiklind \& Combes \cite{wc99}; Bradford et al. \cite{bradford05}), the fractional 
abundance of NH$_3$ becomes $X$(NH$_3$) $\sim$ 3$\times$10$^{-9}$ $\times$ $T_{\rm ex}$/K. For comparison, in nearby cold dark 
clouds the ammonia abundance relative to H$_2$ is $X$(NH$_3$) = 10$^{-7.0...-7.5}$ (e.g., Benson \& Myers \cite{benson83}; 
Hotzel et al. \cite{hotzel04}). In hot cores, that is in regions with significant dust-grain evaporation, abundances can be as 
high as $X$(NH$_3$) = 10$^{-5...-6}$ (e.g., Henkel et al. \cite{henkel87}).

To reach the range of abundances observed in Galactic cloud cores, $T_{\rm ex}$ must be $\ga$10\,K for ammonia inversion 
lines toward PKS\,1830-211. Using a Large Velocity Gradient code with collision rates of Danby et al. (\cite{danby88}), $T_{\rm kin}$ 
= 80\,K and optically thin lines, such a $T_{\rm ex}$ value is obained at a density of $n$(H$_2)$ $\sim$ 10$^{4}$\,cm$^{-3}$. This 
is much higher than the density obtained by Jethava et al. (\cite{jethava07}), $n$(H$_2$)$<$10$^3$\,cm$^{-3}$ (most likely 200\,cm$^{-3}$) 
toward B0218+357, but is still consistent with the upper limits derived using observations of the rotational lines of linear molecules 
toward PKS\,1830--211. An excitation temperature of $<$15\,K from CO $J$=4--3 and 5--4 (Gerin et al. \cite{gerin97}) yields 
$n$(H$_2$) $<$ 3$\times$10$^{3}$\,cm$^{-3}$. However, CO with its small dipole moment and its high abundance, providing particularly 
effective shielding against incoming radiation, may not trace the gas component seen by NH$_3$. More relevant are upper limits of 
6\,K from high density tracers like CS, HCN, and HCO$^+$ (Wiklind \& Combes \cite{wc96b}; Gerin et al. \cite{gerin97}; Combes \& 
Wiklind \cite{cw99}; Muller et al. \cite{muller06}), that require $n$(H$_2$) $\la$ 10$^{4}$\,cm$^{-3}$. This limits $T_{\rm ex}$(NH$_3$) 
to values $\la$10\,K, while the microwave background at $z$ = 0.886 provides a firm lower limit of 5\,K. With $T_{\rm ex}$(NH$_3$) = 
5--10\,K, the NH$_3$ column density and fractional abundance become $N$(NH$_3$) $\sim$ (5--10)$\times$10$^{14}$\,cm$^{-2}$ and 
$X$(NH$_3$) $\sim$ (1.5--3.0)$\times$10$^{-8}$ toward PKS\,1830-211. $X$(NH$_3$) is close to the lower limit of what has been obtained 
from Galactic cloud cores.

The apparent low fractional abundance of ammonia may be caused (1) by stellar nucleosynthesis, (2) by a diffuse medium of low density 
or (3) by differences in the morphology of the radio continuum background at cm- and mm-wavelengths. In the following, these 
possibilities are discussed.

(1) Ammonia is a nitrogen-bearing molecule. $^{14}$N is a mostly ``secondary'' nucleus that should be underabundant relative
to carbon and even more so relative to oxygen in ``chemically'' rather unprocessed spiral galaxies in the distant past (see
Wheeler et al. \cite{wheeler89} for a review on metallicity dependent CNO abundances). A small metallicity also reduces the 
dust-to-gas mass ratio so that shielding by dust grains becomes less effective. Since NH$_3$ is a molecule which is particularly
sensitive to photodissociation (e.g., Wei{\ss} et al. \cite{weiss01}), this could deplete ammonia below values commonly found 
in local molecular clouds.

(2) In Galactic diffuse clouds seen in absorption toward background continuum sources, Liszt et al. (\cite{liszt06}) find that 
the NH$_3$ column density is well correlated only with those of CS and H$_2$CO. NH$_3$/CS and NH$_3$/H$_2$CO abundance ratios 
are approximately 1.0 and 0.4. For PKS\,1830-211, we obtain, using the column densities of Menten et al. (\cite{menten99}), ratios of 
1.0--2.0 and 1.3--2.6. This is approximately consistent, accounting for uncertainties, the small number of studied Galactic clouds, 
and their scatter with respect to relative abundances. Adopting, as for PKS\,1830--211, a ratio $N$(H$_2$)/$N$(CO) $\sim$10$^4$, the 
ammonia abundances for Galactic diffuse clouds seen in absorption become $N$(NH$_3$)/$N$(H$_2$) $\sim$ (1--7)$\sim$10$^{-8}$ (Liszt \& Lucas 
\cite{liszt98}; Liszt et al. \cite{liszt06}), again in reasonable agreement with our result for PKS\,1830--211. 

(3) The third explanation requires that the size of the continuum background source be larger at cm- than at mm-wavelengths.
In the case of B0218+357, this affects the molecular column densities averaged over the continuum source so that optical depths of
ammonia and formaldehyde from lines at low frequencies ($<$30\,GHz) have to be corrected upwards when being compared with lines
observed in the 1--3\,mm wavelength range (Henkel et al. \cite{henkel05}; Jethava et al. \cite{jethava07}). It is possible
that such a behavior also affects the opacities measured toward PKS1830-211.

While the relative importance of the three scenarios explaining a low NH$_3$ abundance cannot be evaluated, we note that the 
NH$_3$ abundance is not enhanced by grain mantle evaporation, which is common in dense and warm Galactic hot cores and which
was suggested by Muller et al. (\cite{muller06}) for H$_2$S toward PKS\,1830-211. 

If we adopt an H{\sc i} spin temperature of 80\,K, valid for the bulk of the ammonia absorption (Sect.\,4.2), the total
background continuum averaged H{\sc i} column density would be 1.6$\times$10$^{21}$\,cm$^{-2}$ (see Koopmans \& de Bruyn
\cite{koopmans05}). In this case, the line-of-sight is predominantly molecular toward the south-western continuum hotspot of
PKS\,1830-211. If we take 600\,K, however, H{\sc i} and H$_2$ column densities would be similar.

\subsection{An analog in the local Universe}

In the nearby Universe, there is only one known cloud that shows a series of metastable NH$_3$ absorption lines similar to that 
seen toward PKS1830--211. It is observed along the line-of-sight toward the prominent star forming-region Sgr~B2 close to 
the center of the Galaxy (Wilson et al. \cite{wilson82}; H{\"u}ttemeister et al. \cite{huette93}, \cite{huette95}; Ceccarelli
et al. \cite{ceccarelli02}). Wilson et al. (\cite{wilson06}) even detected the ($J$,$K$) = (18,18) transition, 3130\,K above 
the ground state, toward Sgr~B2\,(M) and likely also toward Sgr~B2\,(N). Radial velocity and excitation clearly indicate 
that the absorbing component is not a foreground object from the Galactic disk but is associated with Sgr~B2 itself.

Two scenarios were proposed to explain the warm (up to $>$1000\,K) molecular gas seen in absorption toward Sgr~B2. These imply
either subthermally excited gas of low density from the extended envelope, known to surround the star-forming region, or thermalized
dense gas from compact molecular hot spots in the immediate vicinity of the numerous H{\sc ii} regions of Sgr~B2. The low density 
of the gas found toward PKS\,1830-211 suggests by analogy that the Sgr~B2 gas component is also diffuse.

Wilson et al. (\cite{wilson06}) argued that the warm, extended molecular envelope of low density, surrounding Sgr~B2 and being seen
in ammonia absorption up to at least the ($J$,$K$) = (7,7) doublet, may be caused by the peculiar physical conditions close to the Galactic
center. A large stellar density as well as a high probability of cloud-cloud collisions may provide sufficient molecular debris to create
such an environment with higher likelihood than in Galactic disks. In view of this it is noteworthy that, in the $z$=0.9 lensing
galaxy of PKS\,1830--211, we have found a region with similar properties (but apparently lacking a dense molecular core) at a 
galactocentric radius of $\sim$4\,kpc.

\section{Conclusions and outlook}

Our ammonia observations of the gravitational lens system PKS\,1830--211 at a redshift of $z$$\sim$0.9 reveal the following main
results:

\begin{itemize}

\item Ammonia (NH$_3$) is detected in absorption in its ten lowest metatstable inversion lines. The ($J$,$K$) = (10,10)
transition lies $\sim$1030\,K above the ground state.

\item The NH$_3$ absorption lines are optically thin and reach, in the strongest line, only 2.5\% of the continuum level.
No ammonia absorption is detected toward the system displaced by \hbox{--147}\,km\,s$^{-1}$ relative to the main component.

\item The absorbing gas is warm: 80-90\% of the absorption arises from an environment with $T_{\rm kin}$ $\sim$ 80\,K. The
remaining gas is with $>$100\,K considerably warmer; for some gas, $T_{\rm kin}$$\ga$600\,K. The total column density
is approximately $N$(NH$_3$) = (5--10)$\times$10$^{14}$\,cm$^{-2}$.

\item Compared with H$_2$ column densities derived from CO at mm-wavelengths, the fractional abundance is $X$(NH$_3$) $\sim$ 
1.5--3.0$\times$10$^{-8}$. This is a low value when compared with abundances in Galactic cloud cores. The low value may be caused by a 
low metallicity implying small abundances of nitrogen and dust (NH$_3$ is easily destroyed by UV-photons), by diffuse gas that also 
shows low NH$_3$ abundances along Galactic lines-of-sight or it may reflect differences between the continuum background 
morphology at cm- and mm-wavelengths.   

\item If the H{\sc i} spin temperature is equal to the kinetic temperature of the bulk of the NH$_3$ absorbing gas (80\,K),
the total column must be mostly molecular. If, however, most of the H{\sc i} is associated with the hot ammonia component, 
the H{\sc i} and molecular column densities should be similar.

\item In the nearby Universe, only the envelope of Sgr~B2 has a cloud component with ammonia properties similar to those seen toward
PKS\,1830--211. This cloud contains gas from either hot cores or from low-density gas forming an envelope; the analogy 
with PKS1830--211 suggests the latter. The column toward PKS\,1830--211 is unique in the sense that it arises from a spiral arm 
and not from the central region of a gas-rich galaxy.

\end{itemize}

Following Flambaum \& Kozlov (\cite{flambaum07}), a detailed comparison of NH$_3$ inversion lines with rotational spectra
from linear molecules can provide limits on the space-time variation of the ratio of the proton to electron mass. Making use of 
the ammonia profiles presented here, such a comparison will be carried out in a forthcoming paper.

\begin{acknowledgements}
We wish to thank C. M. Walmsley and T.L. Wilson for critically reading the manuscript.
\end{acknowledgements}

\end{document}